\documentclass[conference]{IEEEtran}
\IEEEoverridecommandlockouts
\usepackage{cite}
\usepackage{amsmath,amssymb,amsfonts}
\usepackage{algorithmic}
\usepackage{graphicx}
\usepackage{subcaption}
\usepackage{textcomp}
\usepackage{xcolor}
\def\BibTeX{{\rm B\kern-.05em{\sc i\kern-.025em b}\kern-.08em
    T\kern-.1667em\lower.7ex\hbox{E}\kern-.125emX}}
\usepackage{tikz}
\usetikzlibrary{positioning}

\begin{document}

\title{
Recommendations for Verifying HDR Subjective Testing Workflows
\thanks{This work was funded by DTIF EI Grant No DT-2019-0068 and The ADAPT SFI Research Center.\newline\newline}
}

\author{\IEEEauthorblockN{
Vibhoothi$^{\dagger}$, 
Angeliki Katsenou$^{\dagger\ddagger}$, 
John Squires$^{\dagger}$, 
Fran\c{c}ois Piti\'e$^{\dagger}$, Anil Kokaram$^{\dagger}$}
\IEEEauthorblockA{
$^{\dagger}$ Sigmedia Group, Department of Electronic and Electrical Engineering, \textit{Trinity College Dublin}, Ireland \\
$^{\ddagger}$  Department of Electrical and Electronic Engineering, \textit{University of Bristol}, United Kingdom \\
$^{\dagger}$\{vibhoothi, john.squires, pitief, anil.kokaram\}@tcd.ie, $^{\ddagger}$\{angeliki.katsenou\}@bristol.ac.uk}
}

\maketitle

\IEEEoverridecommandlockouts
\IEEEpubid{
\makebox[\columnwidth]{979-8-3503-1173-0/223/\$31.00
\copyright 2023 IEEE \hfill
} 
\hspace{\columnsep}\makebox[\columnwidth]{}
}

\begin{abstract}
Over the past few years, there has been an increase in the demand and availability of High Dynamic Range (HDR) displays and content. To ensure the production of high-quality materials, human evaluation is required. However, ascertaining whether the full playback pipeline is indeed HDR-compliant can be challenging. 
In this paper, we present a set of recommendations for conformance testing to validate various aspects of the testing workflow, including playback, displays, brightness, colours, and viewing environment. We assessed the effectiveness of HDR conversion techniques used in current standards development (3GPP) for making source materials. Additionally, we evaluate HDR display technologies, including OLED and LCD, using both consumer television and a reference monitor.

\end{abstract}

\begin{IEEEkeywords}
HDR, testing workflow, testing environment, playback, video coding.
\vspace{-1em}
\end{IEEEkeywords}
\begin{tikzpicture}[overlay, remember picture]
\path (current page.north) node (anchor) {};
\node [below=of anchor] {%
2023 15th International Conference on Quality of Multimedia Experience (QoMEX)};
\end{tikzpicture}

\section{Introduction}
In recent years, HDR displays sales and content delivery have increased~\cite{2022smpteuhdprogress,2021-hdrprogres-smptedobly}. Oftentimes, the displayed video output may not always be true HDR. It is hard to confirm whether the playback is actually HDR or to cross-check a playback pipeline disruption. Validating this is important for quality assessment studies and it yields technical challenges. 

The development of standards testing protocols for subjective viewing environments for HDR videos is being led by the Video Quality Experts Group (VQEG), International Telecommunication Union (ITU) Recommendations, and the European Broadcasters Union (EBU) for both televisions and broadcast applications~\cite{2021-bt2390-hdr-tvprod, 2019-t3320-hdr-requirements, 2019-tr47-ebu-hdrmonitor, 2022-t3325-hdr-measurement}. Despite efforts made by previous authors to explore and adapt these techniques for HDR subjective studies\cite{2016hdrsubjectivehevc, 2019hdrhevcsubjective, 2022hdrambient,2023hdrsportshevc},
there are still issues unresolved.
Particularly, the configuration and technical validation of the playback pipeline is often not transparently presented in the literature. Thus, there is currently a lack of an HDR quality assessment framework that can be readily used to conform with these standards.

Taking into account all the above, in this paper, we introduce an HDR quality assessment testing workflow (Section~\ref{sec:validating-playback}), which comprises three important elements, a) the documentation of the playback pipeline, b) the HDR intermediate file conversions, and c) the testing environment.



\section{HDR Standards}
\label{sec:background}

The current HDR standard~\cite{itu_hdr} for television and broadcast deployment of HDR content within the streaming media industry is hindered by the fact that this content requires a very-high bitrate for visual retention of cinematic and artistic intent in motion pictures. 
Since 2017, EBU's Video System and Workflows working group have laid out various testing methodologies for HDR picture monitors. 
In 2019, EBU Tech Report TR047~\cite{2019-tr47-ebu-hdrmonitor} and 3320v4.1~\cite{2019-t3320-hdr-requirements} showcased the performance and recommendations for using first-generation HDR studio monitors.
In 2022, Tech Report 3225v2~\cite{2022-t3325-hdr-measurement} expanded with more testing guidelines using specific HDR test patterns.
In a similar timeline (2021), ITU published BT.2390-10~\cite{2021-bt2390-hdr-tvprod}, which explains Perceptual Quantizer (PQ) HDR TV architecture from content acquisition to display.

Given the fact that current ITU subjective testing methodologies~\cite{itubt50014, 2022-itu-p910} do not accurately account for HDR characteristics, different authors designed different methodologies for testing the quality of HDR videos~\cite{2016hdrsubjectivehevc, 2019hdrhevcsubjective, 2022hdrambient,2023hdrsportshevc}. Due to different test mechanisms and different interactions within the testing pipeline, it is unclear whether we can compare them.

To wrap up, there are currently ongoing efforts from various standardisation bodies and researchers to define a methodology for the testing workflow of HDR videos. 

\section{HDR Subjective Testing Workflow}
\label{sec:validating-playback}
To ensure conformity with the modern HDR standard (ITU BT.2100~\cite{itu_hdr}), we require to validate multiple factors for the HDR quality assessment framework. The framework consists of three distinct parts. The first part is for the playback pipeline which includes cross-checking the playback, brightness, colour, and bit-depth of the display device. The second part is for handling intermediate file conversions. Finally, the third part concerns the testing environment. 
\subsection{Playback Pipeline}
\begin{figure*}
    \centering
    \includegraphics[width=0.89\linewidth]{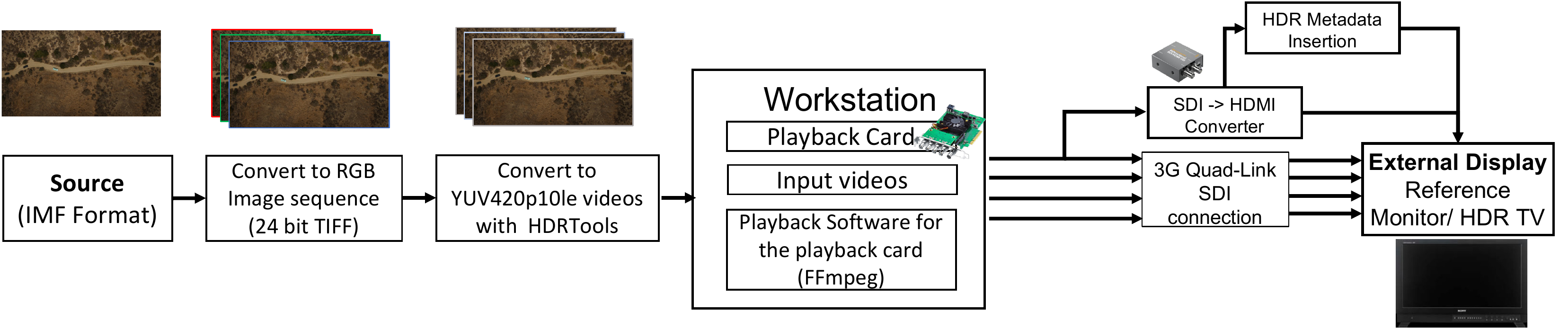}
    \caption{A typical playback pipeline for HDR testing. Input video inside the workstation denotes the final video after conversions.
    }
    \vspace{-.5em}
    \label{fig:workflow-diagram}
\end{figure*}

\subsubsection{Playback}
\label{sec:val:playback}

Figure~\ref{fig:workflow-diagram} outlines the typical playback pipeline to be used in a testing workflow. The initial part of the workflow is conversion and making the source into an encoder-friendly format (More in Section~\ref{sec:hdr-conversion}).
When it comes to HDR video playback, many software video players across different operating systems (OS) do not support true-HDR playback. This is either due to the limitation of hardware or software support (lack of implementation or OS-level support). 
To circumvent this problem, we recommend using dedicated hardware for video playback. In this work, we 
utilised a Blackmagic Decklink 8K Pro Playback device~\cite{2023blackmagic-decklink}
in a Linux environment, where a build of FFmpeg~\cite{2023-ffmpeg-decklink} software with Blackmagic support is used for video playback. Alternatively, the open-source GStreamer, or vendor-specific playback software (e.g. DaVinci Resolve~\cite{2023-davinciresolve} can be utilised.

\vspace{0.25em}
{\noindent \textit{Extension to consumer displays.}}
In modern HDR consumer displays (televisions, or monitors), signalling metadata (colour primaries, transfer characteristics, matrix coefficients etc) is essential for HDR playback. Often, the hardware playback device or any converters which are used in the pipeline would strip the HDR metadata which can result in SDR playback. We recommend forcing HDR metadata on the device end, in cases where it is not available, an intermediate device that inserts HDR metadata is advised (e.g. Dr HDMI from HDFury~\cite{2023-hdfury}).

\vspace{0.25em}
{\noindent \textit{Signal Validation.}} When multiple sets of hardware devices (including various cables) are used in the playback pipeline, signal integrity (or statistics/existence) should be checked. To this end, and for signal passthrough, we recommend using a cross-converter/waveform monitor (we used Atmos Shogun 7).

\subsubsection{Displays}
\label{sec:val:display}
The next milestone to accomplish true HDR video playback is the reliability of the television/monitor's display panel in use. For this, at least five aspects should be observed:
i) the ability to programmatically set the HDR settings in the display device,
ii) option to turn off vendor-specific features for picture quality enhancement (tone-mapping, auto brightness limiter (ABL), gradation etc)
iii) faithful tracking of the electro-optical transfer function (EOTF) in use (PQ) for both low and high-luminance areas,
iv) the ability to display at least 1000 nits of brightness for at least 5-10\% window, 
v) Behaviour of sustained brightness over the period. 
Keeping all of these in consideration, we are utilising a Sony BVM-X300v2 OLED critical reference monitor as a source of reference, along with two consumer-level LCD (Sony KD-75ZD9) and OLED (Sony A80J) HDR display televisions.

\begin{figure}[t]
    \centering
    \includegraphics[width=0.30\linewidth]{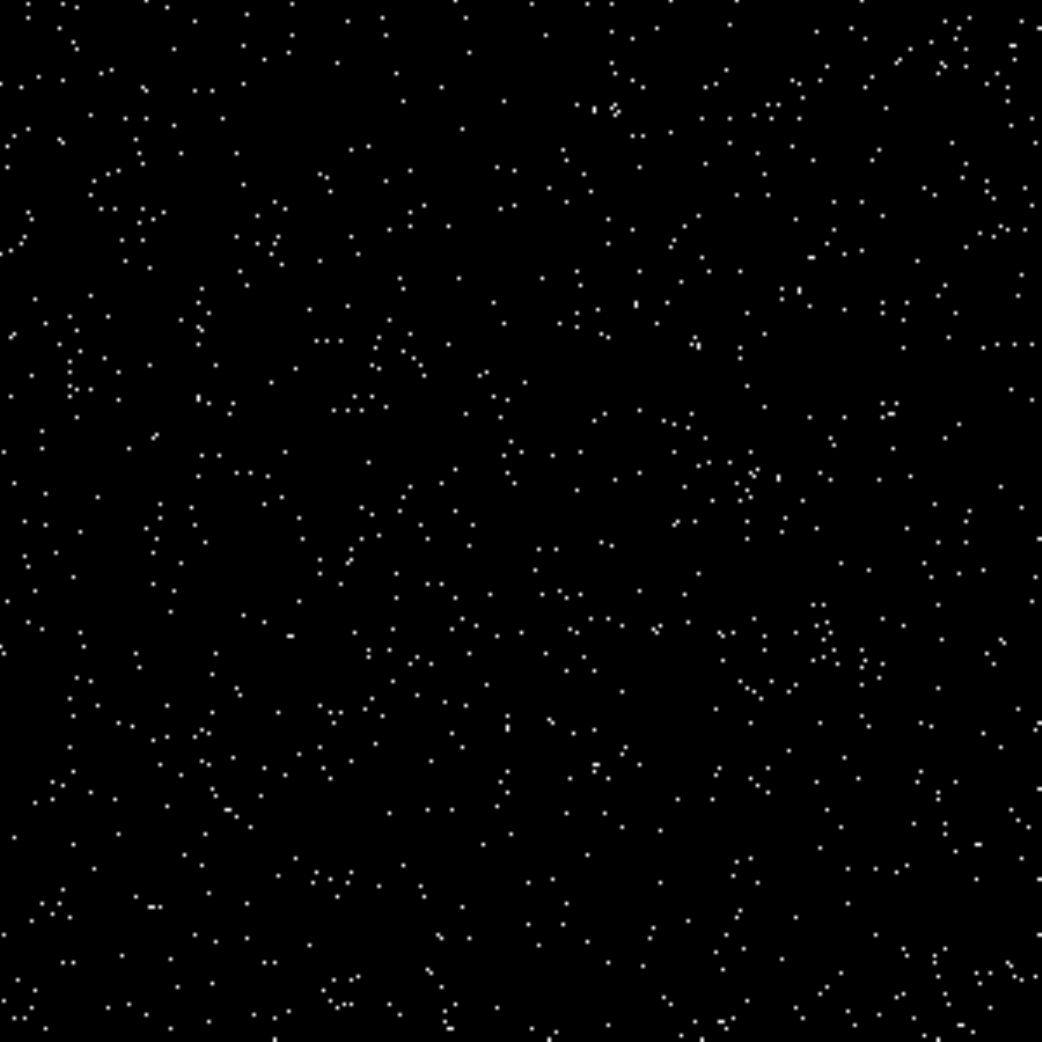}
    \includegraphics[width=0.30\linewidth]{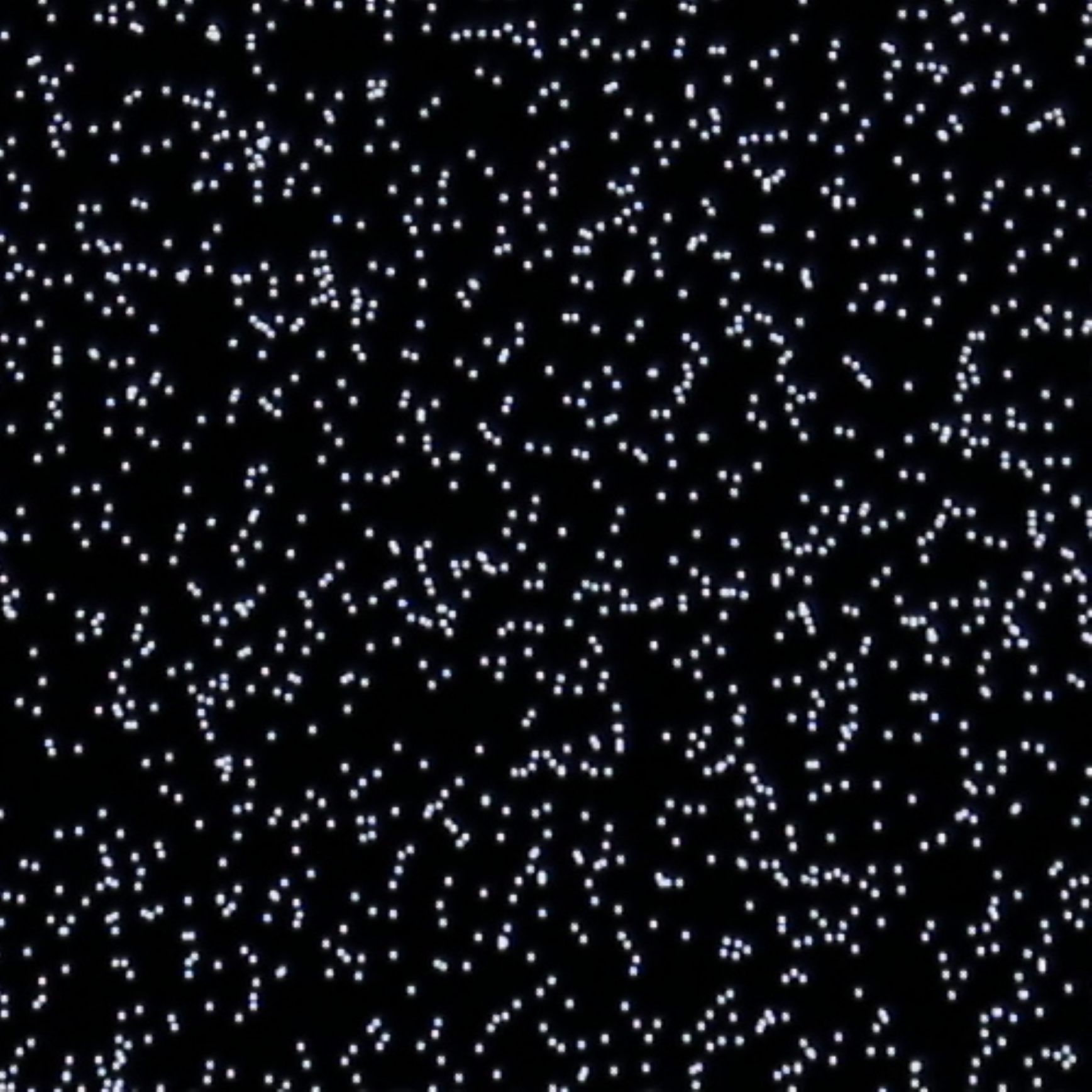}
    \includegraphics[width=0.30\linewidth]{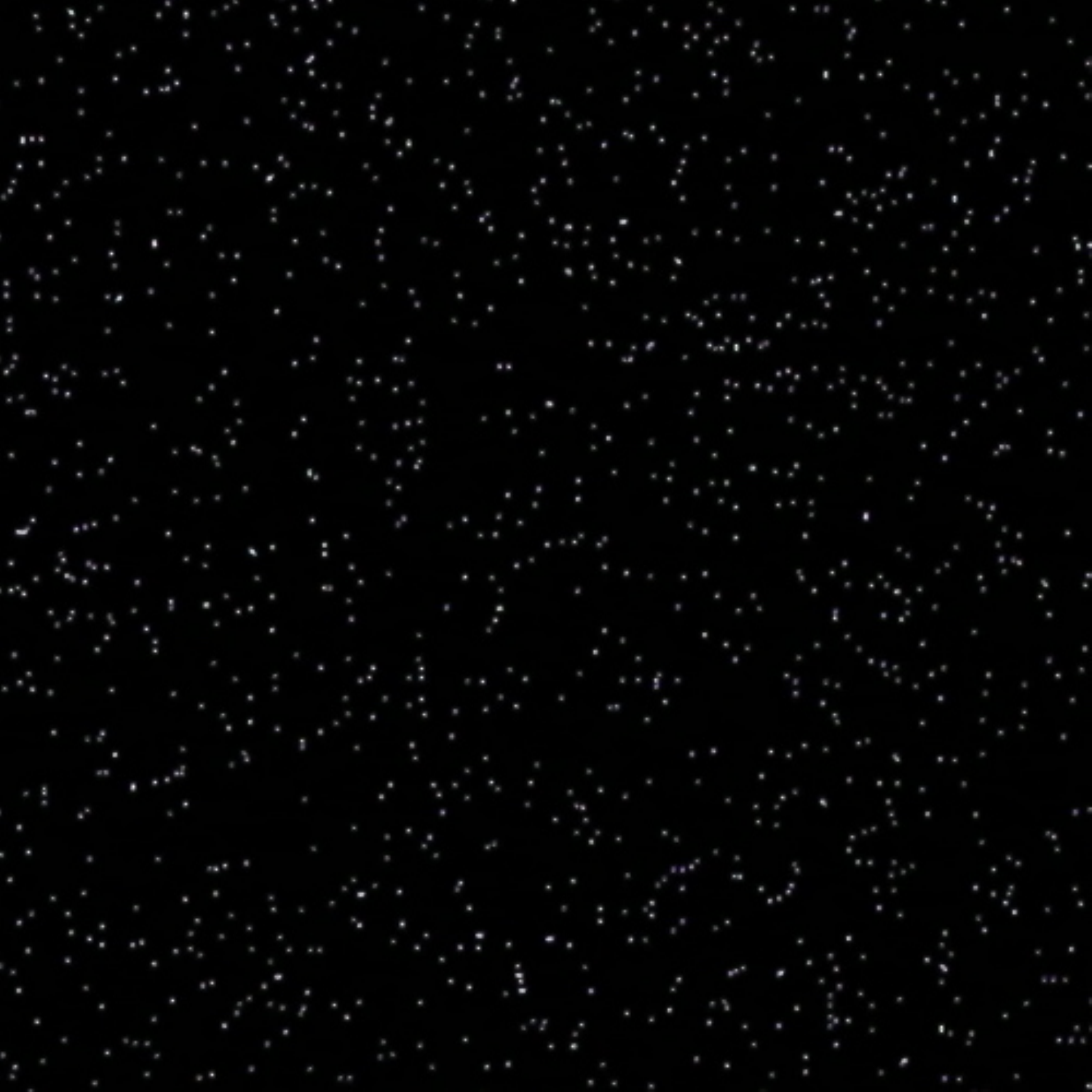}
    \caption{Night-sky test pattern (1\% white-window) across two different display technology, Sony-BVM-X300-V2 (middle) reference monitor, and Sony KD-75ZD9 LCD display (right). The input image is on the left. In the LCD display, white pixels are crushed due to the prominence of black pixels.}
    \vspace{-1.8em}
    \label{fig:night-sky-pattern}
\end{figure}

\vspace{0.25em}
{\noindent \textit{Local dimming analysis.}} To analyse the display panel's local dimming, blooming effect, and colour-bleeding artefacts, we developed a night-sky-star test pattern~\cite{2023qomex-vib-page}. This pattern randomly distributes different percentages (1, 2, 5, 10, 20, 50, 80\%) of peak-white pixels across the display resolution. Figure~\ref{fig:night-sky-pattern} showcases the behaviour of a 1\% white window with a reference monitor (middle) and the Sony LCD TV (right).  
We advise using this artificial test pattern for measuring the true behaviour of the panel over real night-sky patterns as they are prone to ISO camera noise. We later measured the brightness of a small area where most pixels are i) black, and ii) white. 
If a significant increase of brightness over window size for both is observed, the panel is susceptible to poor local dimming. In our study, we observed the brightness of the LCD panel increased linearly based on the number of white pixels, and the OLED panel showcased superior local dimming. 

\subsubsection{Brightness}
\label{sec:val:brightness}
\begin{figure}
    \centering
    \begin{subfigure}[b]{0.46\linewidth}
        \centering
        \includegraphics[width=\linewidth]{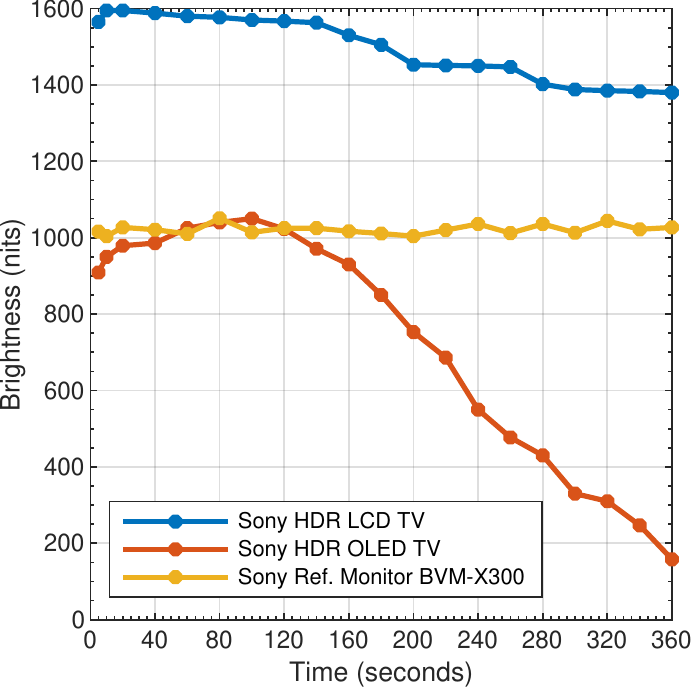}
        \caption{Sustained Brightness variation for different displays}
        \label{fig:sustained-brightness}
    \end{subfigure}
    \begin{subfigure}[b]{0.46\linewidth}
        \centering
        \includegraphics[width=\linewidth]{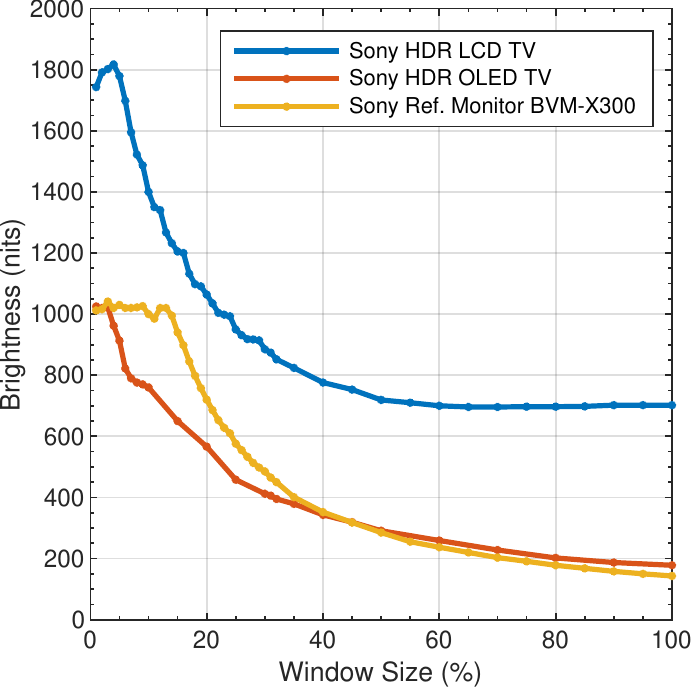}
        \caption{Brightness variation for different window sizes}
        \label{fig:brightness-variation}
    \end{subfigure}
    \caption{Display characterisation for different panels, a reference monitor (yellow), an LCD TV (blue) and OLED TV (red).}
    \vspace{-1.2em}
    \label{fig:display-analysis}
\end{figure}

Many of the current consumer displays have ABLs that do not allow peak brightness beyond a certain window size (gradually decreases) and/or sustained peak brightness over time. Many displays include this to protect the display units. Thus, we recommend analysing i) the sustained brightness using a 1\% full white window; ii) the brightness variation over different window-size.  In EBU's Tech Report 3225v2~\cite{2022-t3325-hdr-measurement}, it is recommended to test the peak brightness ($L$) of the TVs using a full-white window at four levels of screen area ($S$) (4, 10, 25 and 81\%). In our analysis, we discovered that four points may not be sufficient to model the true behaviour of consumer displays. We recommend expanding this by including more steps $S \in \{1\ldots5, 7, 10, 12, 15, 20, 25, 30, 40, 50, 60, 75, 80, 90, 100\}$\%.
Figure~\ref{fig:sustained-brightness} shows the sustained brightness of 1\% window observed for a period of 600 seconds for the three considered display units. As easily observed, the reference monitor (yellow line) consistently sustains the brightness. The LCD TV (blue line) sustained high brightness for a long period ($\approx$180 sec for $>=$1500nits). The  OLED TV (red line) demonstrated a significant drop in brightness after 100 secs. We believe the primary reason for this behaviour is due to heating and the limited cooling of the OLED panel. When the temperature of the TV panel reaches $\approx$55{$^\circ$}C, the peak brightness is obtained (1050 nits), and then the brightness starts quickly decreasing.
Figure~\ref{fig:brightness-variation} shows the variation of brightness of the TVs for increasing window sizes. The Observed peak brightness of the reference monitor was 1041 nits (up to 1000 nits for 13\% window size). The LCD  TV was 1817 nits (up to 1500 nits for 22\% window size). The OLED TV was 1050 nits (up to 900 nits for 5\% window). Both LCD and reference monitors had a smooth degradation of brightness over the growing window size. The OLED TV brightness is very inconsistent due to the heating of the panel, thus for reliability of measurement, we recommend having a cool-off period and monitoring of temperature.

\subsubsection{Colour}
\label{sec:val:colour}
One of the properties of UHD HDR is the availability of wide-colour-gamut (WCG). To ensure compliance with standards, it is necessary to verify the display and signal meet the standard. This can be done through various methods, such as the use of the ``Gamut Marker'' feature in the reference monitor to identify pixels beyond the target colourspace (BT.709/DCI-P3). In other cases, a Spectroradiometer/Colourimeter can precisely measure the wavelength of RGB lights and CIE1931 Chromacity distribution. (we used JETI Spectravel 1511 VIS Spectroradiometer).

\subsubsection{Bit-depth}
\label{sec:val:bitdepth}
It is observed that certain parts of the playback pipeline (at any intermediate step) could decimate some bits (conversion of 10bit to 8bit) and yet have the final playback at 10bit. This can happen either on the playback device side (HDMI modes) or on the display side (pixel format in use). Despite a potential reduction in bit-depth, this is often undetected in playback due to the presence of noise or film grain in the content resulting in smooth gradation. Thus, a fidelity check for bit-depth is recommended. This can be implemented with a grey ramp within the maximum HDR window size of the TV (eg. 5/10\% window size) with 1024 levels/bands/ramps. If a smooth ramp is observed, there is probably no decimation in the playback pipeline (``clean chain''). In all other cases, it denotes a loss of information in the pipeline (``non-pristine chain''). If the input signal contains noise (most of the real-world content), the ``non-pristine chain'', can behave as the ``clean chain''. We cross-checked this, and we observed smooth ramps without banding (same as a ``clean chain'') for a noisy signal. This indicates that HDR fidelity relies on testing materials.

\vspace{-0.3em}
\subsection{Handling HDR Intermediate conversions}
\label{sec:hdr-conversion}
Most modern cameras shoot images and videos in a colour-coded luminance channel, which is later converted to RGB space (de-bayering), and later to an uncompressed intermediate format (IMF format) in video production. The IMF format may not be directly compatible with any given encoder for compression applications. Thus, we require conversion of the videos to Y'CbCr colour space (at 4:2:0 TV range) without losing picture fidelity. This requires multiple visual inspections. 
In 2022, the 3GPP standards body~\cite{2022-tr-3gpp-26.955-av1}, outlined steps taken for the conversion of HDR source videos (Netflix Open-Content's~\cite{netflixopencontent} Sparks, Meridian, Cosmos) from an IMF format (J2K) to an encoder-friendly format using HDRTools~\cite{hdrtools}. 
We tested the conversions using HDRTools with different HDR materials, i) the American Society of Cinematographers' StEM2 (Standard Evaluation Material 2)~\cite{2022-smpte-stem2-paper}, and ii) SVT Open-content~\cite{svt2022}. Later, a cross-check with the original source for colour fidelity using a spectroradiometer was carried out. We observed close reproduction of source information. For a sanity check, a few samples were tested with x265, libaom-av1, and SVT-AV1, and compression and playback were as expected. Thus, we recommend using HDRTools for conversions of HDR materials.

\vspace{-0.2em}
\subsection{Testing environment}
\label{sec:subj-testing-env}
In an HDR subjective testing workflow, 
the viewing environment plays a significant role in the perception of quality~\cite{2009-hdr-viewingenvironment} along with the playback. We recommend validating the following elemental factors: i) the display panel technology (see ~\ref{sec:val:display}), ii) the surrounding environment, light, and reflections from/on the display, iii) the test video content.

Regarding the interface of the subjective study, grey intermediate screens between the display of videos preferred~\cite{2022-itu-p910} to reduce viewing discomfort. The brightness of the grey screen should be configured based on the environment's lighting conditions, video materials in use, and display capabilities. For our experiments, we empirically chose a grey screen (Hex colour code, \textit{\#555555}) of brightness 14.9 nits ($cd/m^2$). Depending on the viewing environment, and excessive exposure to HDR materials, viewers can experience fatigue and dizziness, so it is advisable to have big breaks between viewing sessions. 


\section{Conclusion}
\label{sec:conclusion}
In this work, we proposed and verified various aspects of an HDR subjective testing workflow. This includes the evaluation of various factors such as brightness, colour, display, signal, viewing conditions, and HDR intermediate file conversions for encoding. We conducted experiments to investigate the behaviour of different display technologies in terms of sustained brightness and its variability across window sizes. We observed substantial differences in HDR performance among consumer-level TV models that can affect the presentation and perception of HDR content. However, accounting for all these factors in an experimental environment is accompanied by high costs and is a non-trivial process. 

\bibliographystyle{IEEEtran}
\bibliography{refs}

\end{document}